\documentclass[aps,prb,reprint,amsmath,amssymb]{revtex4-2}
\usepackage{graphicx}
\usepackage{svg}
\usepackage{dcolumn}
\usepackage{bm}
\usepackage{ulem}
\usepackage{soul}
\usepackage{color}
\usepackage{array}
\usepackage{booktabs}
\usepackage{multirow, makecell}
\usepackage{physics}
\usepackage[colorlinks=true, allcolors=blue]{hyperref}%

\begin{document}

\title{Effects of self-consistent extended Hubbard interactions and spin-orbit couplings on energy bands of semiconductors and topological insulators}

\author{Wooil Yang}
\author{Young-Woo Son}
\email{Email: hand@kias.re.kr}
\affiliation{Korea Institute for Advanced Study, Seoul 02455, Korea}

\date{\today}

\begin{abstract}
A first-principles computational method with self-consistent on-site and inter-site Hubbard functionals is able to treat local and non-local Coulomb interactions on an equal footing. To apply the method to understand solids with strong spin-orbit coupling (SOC), we have extended a psuedohybrid functional approach developed by Agapito-Curtarolo-Buongiorno Nardelli to implement self-consistent extended Hubbard energy functionals for noncollinear spin states. With this, energy bands of semiconductors with various SOC strengths such as Si, Ge, GaAs, GaSb, CdSe and PdO are obtained, agreeing with results from fully relativistic $GW$ approximation (FR-GWA) as well as experiments. We also compute energy gaps of HgTe, CuTlS$_2$, and CuTlSe$_2$ and assign them to be topological insulators correctly, unlike characteristic failures for judging topological properties from typical hybrid functionals. We demonstrate feasibility of our method to handle large systems by computing surface bands of topological insulators, Bi$_{\text{2}}$Se$_{\text{3}}$ and Bi$_2$Te$_3$ with varying thickness up to eight quintuple layers. Considering its low computational cost comparable to conventional {\it ab intio} methods and improved accuracy to FR-GWA, we expect that our method provides an opportunity to study large scale correlated systems with the strong SOC efficiently and reliably.
\end{abstract}

\maketitle

\section{Introduction\label{sec:intro}}

First-principles calculations based on density functional theory (DFT) have been a highly successful in simulating various properties of materials efficiently~\cite{HK,KS,Ihm1988RPP,Jones2015RMP,Kurt2016Science}. Concomitantly, there has been a significant progress in using DFT to understand electronic properties of systems having heavy elements, in which the relativistic effects such as strong spin-orbit coupling (SOC) should be taken into account. However, DFT calculations based on approximate exchange-correlation (XC) functionals such as local density approximation (LDA)~\cite{KS} or generalized gradient approximation (GGA)~\cite{PBE} have self-interaction error (SIE) related with heavy elements, posing a challenge to capture their electronic properties~\cite{Cococcioni2005PRB, Cohen2008Science}.

Among various methods to overcome such issues, the DFT+$U$ including the local Coulomb repulsion of $U$ is widely employed~\cite{Anisimov1991PRB,Anisimov1993PRB,Dudarev1998PRB,Himmetoglu2014IJQC}. This is based on the Hubbard model since SIEs occur primarily in localized $d$ and $f$ orbitals. Particularly, recent studies demonstrate accurate description of ground electronic states for several materials  with strong local Coulomb repulsions and SOCs such as transition metal oxides~\cite{kim2008prl}, pyrochlore iridates~\cite{wan2011prb}, americium monochalcogenides~\cite{zhang2012science} and osmium-based double perovskites~\cite{Erickson2007prl}. We also note that  recent developments of fully relativistic $GW$ approximations 
(FR-GWA)~\cite{vasp1996PRB,yambo2009CPC,spex2010prb,spex2011prb,fhiaims2015jctc,scherpelz2016jctc,turbomole2019jcp,questall2020cpc,gpaw2021prb,Barker2022PRB} obtained spin-orbit split energy bands of semiconductors in good agreements with experiments. 

Despite these successes of the DFT+$U$ method, it turns out to be insufficient especially in case that orbital hybridization or inter-site Hubbard interaction of $V$ becomes crucial~\cite{Cococcioni2010JPC,Lee2020PRR,Rubio2020PRB, Timrov2021PRB}. Then, the DFT+$U$+$V$ method based on the extended Hubbard model including both $U$ and $V$ has been considered for several materials with negligible SOCs~\cite{Cococcioni2010JPC,Lee2020PRR,Rubio2020PRB, Timrov2021PRB},  demonstrating its improved capabilities in computing electronic and phonon properties of solids~\cite{Cococcioni2010JPC,Lee2020PRR,Rubio2020PRB, Timrov2021PRB,Yang2021PRB,Yang2022JPC,Ricca2020PRR,Timrov2022PRXE,Mahaja2021PRM, Jang2023PRL}. We also note that a previous work~\cite{Gatti2020PRL} using a localized orbital basis set well demonstrated an importance of including both inter-site Hubbard interactions and SOCs for the nodal-line semimetal ZrSiSe. Therefore, we expect that the DFT with both extended Hubbard interactions and SOC could be another efficient computational tool to calculate reliable physical properties of solids having heavy atomic elements {\it ab initio}.

In this paper, we investigate effects of self-consistent extended Hubbard interactions of $U$ and $V$ on energy gaps and spin-orbit splittings of semiconductors and topological insulators (TIs) with various SOC strengths. 
For semiconductors, we compute spin-orbit 
split energy bands of Si, Ge, GaAs, GaSb, CdSe and PdO.
To compute $U$, $V$ and SOC on an equal footing, 
we extend our previous work ~\cite{Lee2020PRR, Yang2021PRB} on 
self-consistent $U$ and $V$ for collinear states to include noncollinear states. 
Particularly, for self-consistent $V$, we apply the pseudohybrid density 
functionals by Agapito-Curtarolo-Buongiorno Nardelli (ACBN0)~\cite{Agapito2015PRX} to compute the inter-site Hubbard parameters~\cite{Lee2020PRR,Rubio2020PRB} for noncollinear spin states with a plane-wave (PW) basis set~\cite{Giannozzi2009JPC,Giannozzi2017JPC}. 
For TIs, we compute inversion energy gaps of HgTe, CuTlS$_2$ and CuTlSe$_2$ and confirm them to be topologically nontrivial. This is in contrast to some erroneous assignments from hybrid functionals~\cite{Vidal2011PRB}.
Finally, we compute evolution of topological surface states of Bi$_{\text{2}}$Se$_{\text{3}}$ and Bi$_{\text{2}}$Te$_{\text{3}}$ as increasing their slab thickness up to eight quintuple layers (QLs), 
demonstrating our capability to handle large systems.
We show that all results obtained by our new method agree with FR-GWA as well as experiments, thus providing an efficient and reliable 
{\it ab initio} method for large scale systems having heavy atoms.

The paper is organized as follows: in Sec.~\ref{sec:DFT_UV}, we present formalism for DFT+$U$+$V$ method with self-consistent $U$ and $V$ for noncollinear states. The detailed computational parameters for the calculations are presented in Sec.~\ref{sec:details}. In Sec.~\ref{sec:comp_rlt}, we show computational results for atomic and electronic structures of representative semiconductors and TIs with various SOCs and for topological surface states of Bi$_2$Se$_3$ and Bi$_2$Te$_3$ with varying their thickness. We conclude in Sec.~\ref{sec:con}.

\section{Formalism \label{sec:DFT_UV}}
 
We briefly review the extended Hubbard functional formalism ~\cite{Cococcioni2010JPC,Lee2020PRR,Rubio2020PRB} with atomic SOCs built on top of noncollinear DFT  using the norm-conserving pseudopotential (PP) and PW basis set.  
To obtain the generalized occupation matrix for Hubbard interactions by projecting PW wavefunctions on atomic states, the projector characterized by the total angular momentum of $j$ would be required. However, we practically adopt the $j$-averaged radial components of atomic wavefunctions from PPs~\cite{hybertsen1986prb2,hemstreet1993prb,hamann2013prb,Garrity2014CMS}
such as,  
\begin{equation}
V({\bf r},{\bf r'})=V_\text{loc}+\sum_l 
\left[
V^\text{SR}_l({\bf r},{\bf r'})+{\bf L}\cdot{\bf S}V^\text{SO}_l({\bf r},{\bf r'})
\right],
\label{ONCV}
\end{equation}
where ${\bf L}$ and ${\bf S}$ are angular and spin momentum, respectively and $l$ an azimuthal quantum number. 
Here, the non-local potential can be split into scalar-relativisitc (SR) and spin-orbit (SO) components as sums and differences of relativistic potentials of $V^\text{Rel}_j({\bf r},{\bf r'})$ with $j=l+\frac{1}{2}\sigma $ ($j>0$ and spin index $\sigma=\pm 1$)
~\cite{hybertsen1986prb2,hemstreet1993prb,hamann2013prb,Garrity2014CMS} such that
\begin{eqnarray}
V^\text{SR}_l &=&\frac{(l+1)V^\text{Rel}_{l+1/2}+lV^\text{Rel}_{l-1/2}}{2l+1}, \nonumber\\
V^\text{SO}_l &=&2\frac{V^\text{Rel}_{l+1/2}-V^\text{Rel}_{l-1/2}}{2l+1}.
\label{psp}
\end{eqnarray}

Although the $j$-averaged atomic projector from potentials in Eqs.~\eqref{ONCV} and~\eqref{psp} is not spin-dependent, we can practically assign spinor representation for the projector following the previous studies~\cite{Rubio2020PRB,binci2023prb}. 
Then, the mean field Hubbard energy (${\mathcal E}_{MF}$) 
including on-site ($U_I$) and inter-site ($V_{IJ}$) electron-electron interactions
for $I$-th and $J$-th atoms are split into collinear (${\mathcal E}^{\text c}_U$ and ${\mathcal E}^{\text c}_V$) and noncollinear (${\mathcal E}^{\text {nc}}_U$ and ${\mathcal E}^{\text {nc}}_V$) contributions such that,
\begin{eqnarray}
{\mathcal E}_{\text{MF}}&=&{\mathcal E}^{\text c}_U+{\mathcal E}^{\text{nc}}_U+{\mathcal E}^{\text c}_V+{\mathcal E}^{\text{nc}}_V\nonumber \\
&=& \sum_I \left[ E^{\text c}_{U_I}+E^{\text{nc}}_{U_I}\right]+
\sum_{\{I,J\}} \left[ E^{\text c}_{V_{IJ}}+
E^{\text{nc}}_{V_{IJ}}\right],
\label{mfeq}
\end{eqnarray}
where $\{I,J\}$ denotes a pair of different atoms $I$ and $J$ within a given cutoff distance.
Each term in Eq.~\eqref{mfeq} can be expressed as
\begin{eqnarray}
E^{\text c}_{U_I} &=&\frac{1}{2} \sum_{ij,\sigma\sigma'}
(Ii\sigma,Ii\sigma|Ij\sigma',Ij\sigma')
n^{II\sigma\sigma}_{ii} n^{II\sigma'\sigma'}_{jj} \nonumber \\
& &-\frac{1}{2}\sum_{ij,\sigma}
(Ii\sigma, Ij\sigma|Ij\sigma,Ii\sigma) n^{II\sigma\sigma}_{ii} n^{II\sigma\sigma}_{jj}, 
\label{ecu}\\
E^{\text{nc}}_{U_I} &=&
-\frac{1}{2}\sum_{ij,\sigma}
(Ii\sigma,Ij{\sigma}|Ij\bar{\sigma}, Ii\bar{\sigma})
n^{II\sigma\bar{\sigma}}_{ii}
n^{II\bar{\sigma}\sigma}_{jj}, 
\label{encu}\\
E^\text{c}_{V_{IJ}} & =&\frac{1}{2}
\sum_{ij,\sigma\sigma'} 
(Ii\sigma,Ii\sigma|Jj\sigma',Jj\sigma')\nonumber \\
& & \times \left[n^{II\sigma\sigma}_{ii}n^{JJ\sigma'\sigma'}_{jj}-
\delta_{\sigma,\sigma'}n^{IJ\sigma\sigma'}_{ij}n^{JI\sigma'\sigma}_{ji}\right], 
\label{ecv} \\
E^{\text{nc}}_{V_{IJ}} & =&-\frac{1}{2}
\sum_{ij,\sigma} 
(Ii\sigma,Ii\sigma|Jj\bar{\sigma}Jj\bar{\sigma})
n^{IJ\sigma\bar{\sigma}}_{ij}n^{JI\bar{\sigma}\sigma}_{ji}.
\label{encv}
\end{eqnarray}
Here,  
the generalized occupation number $n^{IJ\sigma\sigma'}_{ij}$ can be written as 
\begin{eqnarray}
n^{IJ\sigma\sigma'}_{ij} &\equiv&n^{I,n,l,\sigma,J,n',l',\sigma'}_{ij} \nonumber \\  
&=& \sum_{m\mathbf{k}}w_\mathbf{k}f_{m\mathbf{k}} \langle \psi^{\sigma}_{m\mathbf{k}} | \phi^{I,n,l}_{i\sigma} \rangle \langle \phi^{J,n',l'}_{j\sigma'} | \psi^{\sigma'}_{m\mathbf{k}} \rangle,
    \label{gocc}
\end{eqnarray}
where $f_{m\mathbf{k}}$ is the Fermi-Dirac function of the PW Bloch state $| \psi^{\sigma}_{m\mathbf{k}} \rangle$ of the $\textit{m}$-th band at a momentum $\mathbf{k}$ and $\textit{w}_{m\mathbf{k}}$ the $\mathbf{k}$-grid weight. 
$| \phi^{I,n,l}_{i\sigma} \rangle$ is the L\"{o}wdin orthonormalized atomic wavefunction~\cite{Lowdin1950JCP} as a projector where $n$ is the principal quantum numbers.
$\bar{\sigma}$ denotes an opposite spin orientation to $\sigma$.
We will use an abbreviated notation of $I$, which specific principal and azimuthal qunatum numbers of the $I$-th atom hereafter, $e.g.$, $|\phi^I_{i\sigma}\rangle \equiv | \phi^{I,n,l}_{i\sigma} \rangle $. We note that, when $I=J$, Eq.~\eqref{gocc} becomes the on-site occupation, {\it i.e.}, $n^{I}_{i\sigma}\equiv n^{II\sigma\sigma}_{ii}$.
Here, $i,\cdots, l$ represent the magnetic quantum number (note that $l$ is not azimuthal quantum number hereafter).
The Coulomb integrals are written using the notation, 
\begin{eqnarray}
& &(Ii\sigma,Ij\sigma|Jk\sigma',Jl\sigma') 
\nonumber\\
&\equiv&\langle Ii\sigma,Jk\sigma'|V_{ee}|Ij\sigma,Jl\sigma'\rangle 
\nonumber\\
&=&\int d{\bf r}
\int d{\bf r'}
\frac{
\rho_{ij\sigma}({\bf r}-{\bf R}_I)\rho_{kl\sigma'}({\bf r'}-{\bf R}_J)}
{|{\bf r}-{\bf r'}-{\bf R}_I+{\bf R}_J|},
\end{eqnarray}
where
$
\phi_{i\sigma}({\bf r}-{\bf R}_I)=
\langle {\bf r}-{\bf R}_I |
\phi_{i\sigma}^I\rangle
$, $|\phi^I_{i\sigma}\rangle$ denotes an associated atomic orbital state of the $I$-th atom, ${\bf R}_I$ its position, 
$\rho_{ij\sigma}({\bf r}-{\bf R}_I)=
\phi^*_{i\sigma}({\bf r}-{\bf R}_I)
\phi_{j\sigma}({\bf r}-{\bf R}_I)
$ and
$
\rho_{kl\sigma'}({\bf r'}-{\bf R}_J)=
\phi^*_{k\sigma'}({\bf r'}-{\bf R}_J)
\phi_{l\sigma'}({\bf r'}-{\bf R}_J)
$.

Following Dudarev's approach~\cite{Dudarev1998PRB}, we define the averaged on-site 
and inter-site interactions~\cite{Cococcioni2010JPC} such as
\begin{eqnarray}
U_I &=& \frac{\sum_{ij}(Ii\sigma,Ii\sigma|Ij\sigma',Ij\sigma')}{(2l_I+1)^2}, 
\label{ui}\\
J_I &=& \frac{\sum_{i\neq j}(Ii\sigma,Ij\sigma|Ij\sigma,Ii\sigma)}{2l_I(2l_I+1)}, 
\label{ji} \\
V_{IJ} &=&\frac{\sum_{ij}(Ii\sigma,Ii\sigma|Jj\sigma',Jj\sigma')}{(2l_I+1)(2l_J+1)}
\label{vij},
\end{eqnarray}
where $l_I (l_J)$ is an azimuthal quantum number of interested orbitals in the $I(J)$-th atom.
Using Eqs.~\eqref{ui}$\sim$\eqref{vij}, Eqs~\eqref{ecu}$\sim$\eqref{encv} can be rewritten as 
\begin{eqnarray}
E^\text{c}_{U_I} &=& \frac{U_I-J_I}{2}\sum_{i\neq j,\sigma} 
n^{I}_{i\sigma}  n^{I}_{j\sigma} 
+\frac{U_I}{2}\sum_{ij,\sigma}n^{I}_{i\sigma} n^{I}_{j\bar{\sigma}},
\label{eq:ecui}
 \\
E^{\text{nc}}_{U_I} & =& 
- \frac{U_I}{2}\sum_{i,\sigma}n^{I\sigma\bar{\sigma}}_{ii}n^{I\bar{\sigma}\sigma}_{ii}
- \frac{J_I}{2}\sum_{i\neq j,\sigma}n^{I\sigma\bar{\sigma}}_{ii}n^{I\bar{\sigma}\sigma}_{jj},
\label{eq:encui}
\\
E^\text{c}_{V_{IJ}} & =&\frac{V_{IJ}}{2}
\sum_{ij,\sigma\sigma'} 
 \left[n^{I}_{i\sigma}n^{J}_{j\sigma'}-
\delta_{\sigma\sigma'}n^{IJ\sigma\sigma'}_{ij}n^{JI\sigma'\sigma}_{ji}\right], 
\label{eq:ecvi}
\\
E^{\text{nc}}_{V_{IJ}} & =&-\frac{V_{IJ}}{2}
\sum_{ij,\sigma} 
n^{IJ\sigma\bar{\sigma}}_{ij}n^{JI\bar{\sigma}\sigma}_{ji}.
\label{eq:encvi}
\end{eqnarray}

Following Mosey {\it et. al.}~\cite{Mosey2007PRB,Mosey2008JCP} and ACBN0~\cite{Agapito2015PRX, Rubio2017PRB, Lee2020PRR, Rubio2020PRB} approaches, we now rewrite the mean field energies in~Eqs.~\eqref{ecu}$\sim$\eqref{encv} using the noncollinear renormalized density matrices of $P^{IJ\sigma\sigma'}_{ij}$ that can be written as 
\begin{equation}
P^{IJ\sigma\sigma'}_{ij} = \sum_{m\mathbf{k}}w_\mathbf{k}f_{m\mathbf{k}} 
N^{IJ}_{{m\mathbf{k}}} 
\langle \psi^{\sigma}_{m\mathbf{k}} 
| \phi^{I}_{i\sigma} \rangle 
\langle \phi^{J}_{j\sigma'} 
| \psi^{\sigma'}_{m\mathbf{k}} \rangle.
\label{eq:rep}
\end{equation}
Here renormalized occupation number of 
$N^{IJ}_{{m\mathbf{k}}}$ for different atoms ($I\neq J$) is defined as
\begin{equation}
N^{IJ}_{{m\mathbf{k}}} \equiv
\sum_{\{I \}}
{\textrm{tr}}_o ( {\textrm{tr}}_s
(N^{I\sigma\sigma'}_{i,m{\bf k}}))
+
\sum_{\{J \}}
{\textrm{tr}}_o ({\textrm{tr}}_s
(N^{J\sigma\sigma'}_{i,m{\bf k}})),
\label{eq:ren}
\end{equation}
where 
$N^{I\sigma\sigma'}_{i,m{\bf k}}\equiv
\langle \psi^{\sigma}_{m\mathbf{k}} |  \phi^{I}_{i\sigma} \rangle 
\langle \phi^{I}_{i\sigma'} | \psi^{\sigma'}_{m\mathbf{k}} \rangle 
$,
${\textrm{tr}}_{o(s)}$ is a trace over orbitals (spins),
and $\sum_{\{I(J)\}}$ denote summations of all $I(J)$ atoms. For a pair of the same atoms or for the on-site interactions, Eq.~\eqref{eq:ren} reduces to 
$N^{II}_{{m\mathbf{k}}}= \sum_{\{I \}}{\textrm{tr}}_o( {\textrm{tr}}_s
(N^{I\sigma\sigma'}_{i,m{\bf k}}))$. In Eq. \eqref{eq:ren}, we consider traces over spins since $N^{IJ}_{{m\mathbf{k}}}$ plays as a role of screening to charges that should not depend on spins~\cite{Agapito2015PRX, Rubio2020PRB, Lee2020PRR}.
If including off-diagonal spin orientations in Eq.~\eqref{eq:ren}, we confirm erroneous changes of interactions with varying spin orientations. With introduction of Eq.~\eqref{eq:rep}, we rewrite Eqs.~\eqref{ecu}$\sim$\eqref{encv} in most general forms as followings,
\begin{eqnarray}
& &E_{U_I}=\frac{1}{2} \sum_{ijkl,\sigma\sigma'} 
{P}^{I\sigma}_{ij}{P}^{I\sigma'}_{kl}(Ii\sigma,Ij\sigma|Ik\sigma',Il\sigma') 
\label{eq:acbn_u} \\
& &E_{J_I}=-\frac{1}{2}\sum_{ijkl,\sigma\sigma'} 
{P}^{I\sigma\sigma'}_{ij}{P}^{I\sigma'\sigma}_{kl}
(Ii\sigma,Il\sigma|Ik\sigma',Ij\sigma') 
\label{eq:acbn_j} \\
& &E_{V_{IJ}}=\frac{1}{4} \sum_{ijkl,\sigma\sigma'} 
\left[{P}^{I\sigma}_{ij}{P}^{J\sigma'}_{kl}-{P}^{IJ\sigma\sigma'}_{il}{P}^{JI\sigma'\sigma}_{kj}\right] \nonumber\\
& &~~~~~~~~~~~~~~~~~~~~~~
\times(Ii\sigma,Ij\sigma|Jk\sigma',Jl\sigma'),
\label{eq:acbn_v}
\end{eqnarray}
where $P^{I\sigma}_{ij}\equiv P^{II\sigma\sigma}_{ij}$
and $P^{I\sigma\sigma'}_{ij}\equiv P^{II\sigma\sigma'}_{ij}$.
Then, by equating corresponding terms between Eqs.~\eqref{eq:ecui}$\sim$\eqref{eq:encvi} and 
Eqs.~\eqref{eq:acbn_u}$\sim$\eqref{eq:acbn_v}, respectively, we can obtain the pseudohybrid-type extended Hubbard functionals for the noncollinear states~\cite{Agapito2015PRX,Rubio2017PRB,Rubio2020PRB} within 
DFT+$U$+$V$ formalism~\cite{Cococcioni2010JPC,Lee2020PRR},
\begin{eqnarray}
{\mathcal U}_{I}&=&
 \sum_{ijkl,\sigma\sigma'} 
\frac{{ P}^{I\sigma}_{ij}{ P}^{I\sigma'}_{kl}}{{\mathcal N}_{U_I}}
(Ii\sigma,Ij\sigma|Ik\sigma',Il\sigma'),
\label{eq:uf}\\
{\mathcal J}_I&=& \sum_{ijkl,\sigma\sigma'} 
\frac{
{P}^{I\sigma\sigma'}_{ij}{P}^{I\sigma'\sigma}_{kl}}
{{\mathcal N}_{J_I}}
(Ii\sigma,Il\sigma|Ik\sigma',Ij\sigma') ,
\label{eq:jf}\\
{\mathcal V}_{IJ}&=&
\sum_{ijkl,\sigma\sigma'}
\frac{{P}^{I\sigma}_{ij}{P}^{J\sigma'}_{kl}-{P}^{IJ\sigma\sigma'}_{il}{ P}^{JI\sigma'\sigma}_{kj}}{{\mathcal N}_{V_{IJ}}} 
\nonumber\\
& &\times(Ii\sigma,Ij\sigma|Jk\sigma',Jl\sigma'),
\label{eq:vf}
\end{eqnarray}
where 
\begin{eqnarray}
{\mathcal N}_{U_I} &=& \sum_{i\neq j,\sigma}n^I_{i\sigma}n^I_{j\sigma}+\sum_{ij,\sigma}
n^I_{i\sigma}n^I_{j\bar{\sigma}}-\sum_{i,\sigma}n^{I\bar{\sigma}\sigma}_{ii}n^{I\bar{\sigma}\sigma}_{ii},\nonumber\\
{\mathcal N}_{J_I} &=& \sum_{i\neq j,\sigma} 
\left[ n^I_{i\sigma}n^I_{j\sigma}+n^{I\sigma\bar{\sigma}}_{ii} n^{I\bar{\sigma}\sigma}_{jj}
\right], \nonumber \\
{\mathcal N}_{V_{IJ}} &=&\sum_{ij,\sigma\sigma'}
\left[{n^{I\sigma}_{i}n^{J\sigma'}_{j}}-{n^{IJ,\sigma\sigma'}_{ij}n^{JI,\sigma'\sigma}_{ji}}
\right]. \nonumber
\end{eqnarray}

Considering the double counting corrections within the fully localized limit as discussed before~\cite{Liechtenstein1995PRB,Anisimov1993PRB,Czyzyk1994PRB,Solovyev1994PRB, Bultmark2009prb, Cococcioni2010JPC,Ryee2018SR},  
Eqs.~\eqref{eq:ecui}$\sim$\eqref{eq:encvi} have following double counting energies (${\mathcal E}_\text{dc}$) for on-site ($E^\text{dc}_{U_I}$) and inter-site ($E^\text{dc}_{V_{IJ}}$) parts,

\begin{eqnarray}
E^\text{dc}_{U_I} &=& {\frac{U_I}{2}}(N^2_{I}-N_I)-\frac{J_{I}}{2}\left(\frac{N^2_{I}+|{\bf m}_I|^2}{2}-N_I \right)\label{eq:dcu} 
\\
E^\text{dc}_{V_{IJ}}&=&{\frac{V_{IJ}}{2}}N_{I}N_{J} \label{eq:dcv}
\end{eqnarray}
where $N_I={\textrm{tr}}_o ({\textrm{tr}_s} (n^{I\sigma\sigma'}_{ij}))$ and 
${\bf m}_I ={\textrm{tr}}_o ({\textrm{tr}_s} ({\bm{\sigma}}n^{I\sigma\sigma'}_{ij}))$.

After subtracting Eqs.~\eqref{eq:ecui}$\sim$\eqref{eq:encvi} by Eqs.~\eqref{eq:dcu} and \eqref{eq:dcv} and then replacing on-site and inter-site interactions in Eqs.~\eqref{ui},~\eqref{ji} and~\eqref{vij} with Eqs~\eqref{eq:uf},~\eqref{eq:jf} and~\eqref{eq:vf} respectively, the resulting energy functionals with SOC and without double countings can be written as
\begin{eqnarray}
{E}_{U_I}&=&\frac{{\mathcal U}_I-{\mathcal J}_I}{2}
\left[
\sum_{i,\sigma}n^{I}_{i\sigma}
-\sum_{ij,\sigma\sigma'}
{n^{II\sigma\sigma'}_{ij}n^{II\sigma'\sigma}_{ji}}
\right],
\label{eq:ufinal}
\\
E_{V_{IJ}}&=&-{\frac{{\mathcal V}_{IJ}}{2}}\sum_{ij,\sigma\sigma'}
{n^{IJ\sigma\sigma'}_{ij}n^{JI\sigma'\sigma}_{ji}}.
\label{eq:vfinal}
\end{eqnarray}
Compared with earlier results without SOC~\cite{Agapito2015PRX,Rubio2017PRB,Lee2020PRR,Rubio2020PRB}, 
the forms of the functionals are quite similar and have additional summations over all spin orientations owing to noncollinear states.

\section{Computational details\label{sec:details}}

Our {\it ab initio} calculations have been performed with {\sc Quantum Espresso (QE)} package~\cite{Giannozzi2009JPC,Giannozzi2017JPC}. We employed LDA parametrized by Perdew and Zunger (PZ)~\cite{LDApz}, and GGA by Perdew-Burke-Ernzehof (PBE)~\cite{PBE} and its revision for solids (PBEsol)~\cite{PBEsol}. We used the optimized norm-conserving Vanderbilt PPs~\cite{hamann2013prb} with parameters from the Pseudo Dojo database~\cite{Setten2018CPC}. For scalar relativistic (SR) calculations, we include relativstic mass correction and Darwin term together with $V^\text{SR}_l$ in Eq.~\eqref{ONCV}. For fully relativistic (FR) ones, we further add $V^\text{SO}_l$ in Eq.~\eqref{ONCV} on top of the aformentioned ones. Unless otherwise mentioned, we will use FR(SR)-LDA and FR(SR)-GGA for FR (SR) calculations with FR(SR)-PZ and FR(SR)-PBEsol, respectively. Then, unless otherwise mentioned, we call our method as FRL+$U$+$V$ if using FR-PZ approximation for DFT+$U$+$V$ calculation and FRG+$U$+$V$ if using FR-PBEsol. 

We employed our modified in-house QE package to obtain self-consistent extended Hubbard parameters with PW basis sets~\cite{Lee2020PRR} and SOCs. 
We use L\"{o}wdin orthogonalized atomic orbitals for projectors as were done in our previous studies~\cite{Lee2020PRR, Yang2021PRB, Yang2022JPC}. In calculating $U$ and $V$ in Eqs.~\eqref{eq:uf}$\sim$\eqref{eq:vf}, we use PPs without the semicore in valences
because the occupation numbers in valences using 
the $j$-averaged atomic projectors
slightly vary whether the semicore states are included or not. 
Although the difference due to semicore is not significant, to reflect charge densities from semicore states reliably, the Hubbard parameters are first obtained with SR approximations, if we use LDA PPs having semicore states, and then proceed to FR ones with Eqs.~\eqref{eq:ufinal} and~\eqref{eq:vfinal}.

The on-site \textit{U} for all \textit{s} orbitals was set to be zero. The inter-site interactions of $V$ were included up to the second nearest neighbors for Si, Ge, GaAs, GaSb, Bi$_2$Se$_3$, and Bi$_2$Te$_3$ while for CdSe, PdO, HgTe, CuTlS$_2$, and CuTlSe$_2$, ${\mathcal V}_{IJ}$ is considered for the adjacent atoms only, all of which are confirmed to be converged~\cite{Lee2020PRR}. 
Then, we determined the equilibrium lattice constants and atomic positions with converged on-site and inter-site Hubbard interactions~\cite{Yang2021PRB,Timrov2020PRB}.
We use $15\times15\times15$ $k$-point grid of the Brillouin zone (BZ) sampling for Si, Ge, GaAs, GaSb and HgTe, $15\times15\times 10$ for CdSe and PdO, $11\times11\times11$ mesh for CuTlS$_2$ and CuTlSe$_2$, and $11\times11\times1$ mesh for slab models of Bi$_2$Se$_3$ and Bi$_2$Te$_3$, respectively. We set the kinetic energy cutoff to 120 Ry for Si, Ge, HgTe, CuTlS$_2$, CuTlSe$_2$, Bi$_2$Se$_3$, and Bi$_2$Te$_3$, 350 Ry for GaAs and GaSb, 200 Ry for CdSe and 100 Ry for PdO, respectively.

\section{Results\label{sec:comp_rlt}}

\subsection{Si, Ge, GaAs, and GaSb}

\begin{table} [t]
\caption{Calculated $U$ and $V$ (in eV) with FRG+$U$+$V$ for a set of semiconductors. $U_p$, the on-site Hubbard interaction for $p$ orbital and, $V_{ss}$, $V_{sp}$, and $V_{pp}$, the inter-site Hubbard parameters between $s$-$s$, $s$-$p$, and $p$-$p$ orbitals of the adjacent atoms. For $U_p$ for GaX (X$=$As, Sb), the left and right values are for Ga and X $p$-orbitals, repsectively while, for $V_{sp}$, they are between Ga $s(p)$-orbital and X $p(s)$-orbital. }
\label{tabular:hub_para}
\begin{ruledtabular}
\begin{tabular}{c|cccc} 
     & $U_p$      &$V_{ss}$& $V_{sp}$   & $V_{pp}$\\ \hline
Si   & 3.49       & 0.88    & 0.72       & 1.85   \\ 
Ge   & 3.32       & 1.04    & 0.69       & 1.75   \\ 
GaAs & 0.35, 1.96 & 0.84    & 1.30, 0.81 & 1.67   \\ 
GaSb & 0.43, 1.43 & 0.9     & 1.07, 0.79 & 1.43   \\
\end{tabular}
\end{ruledtabular}
\end{table}

\begin{figure}[b]
	\begin{center}
		\includegraphics[width=1\columnwidth]{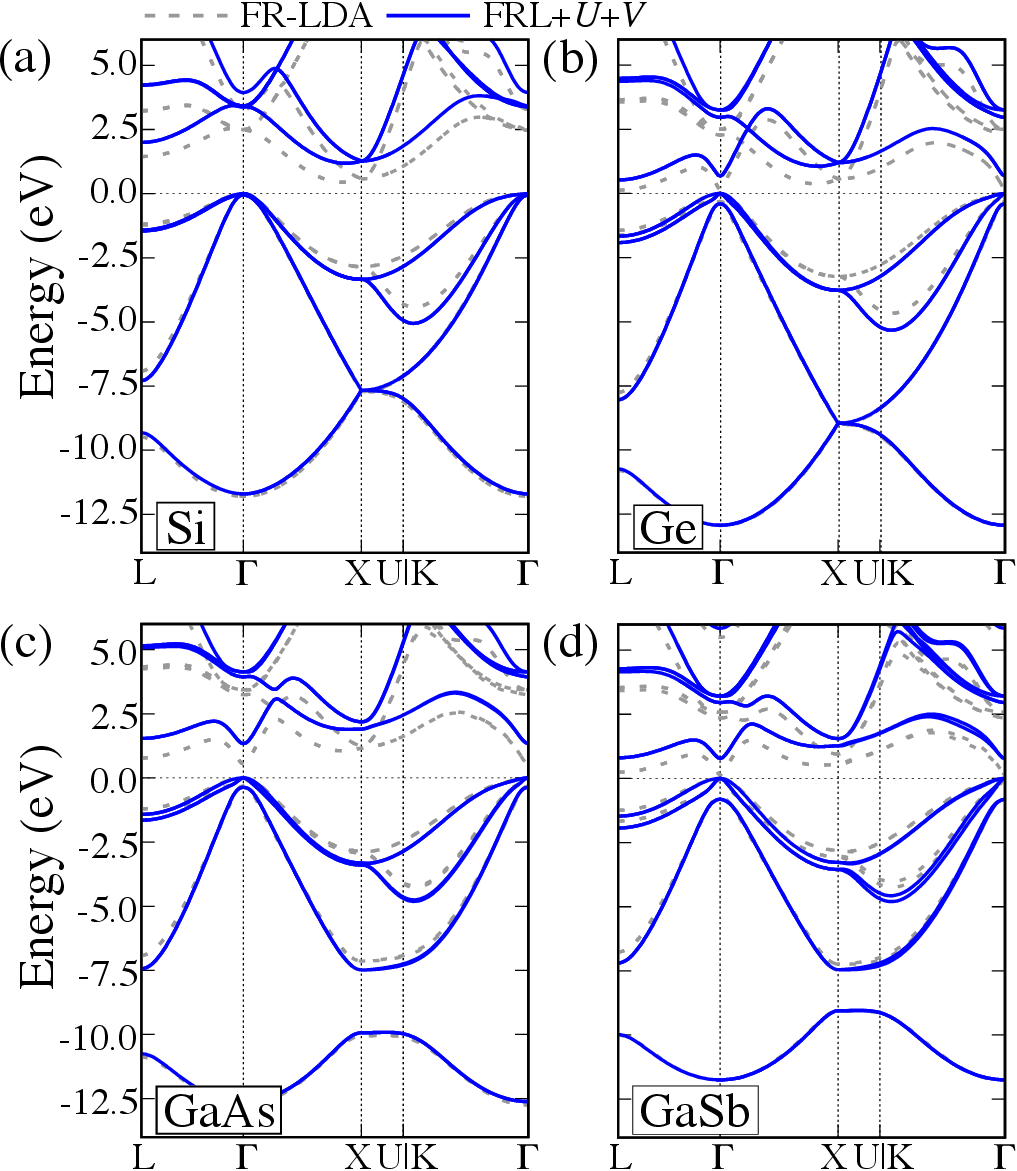}
	\end{center}
	\caption{FR-LDA and FRL+$U$+$V$ band structures of Si, Ge, GaAs, and GaSb, plotted along high-symmetry lines of the Brillouin zone. The valence band maximum is set to zero.
	}
	\label{fig:si}
\end{figure}

We first consider Si, Ge, and GaAs that are prototypical semiconductors to test our method and to compare calculation results with those from other methods as well as many experiments.
We also compute atomic and electronic structures of GaSb of which spin-orbit splitting in its valence bands is comparable to its band gap.

Table \ref{tabular:hub_para} shows the calculated self-consistent $U$ and $V$ values. We have checked that for all of them, which are either diamond or zincblende semiconductors, the calculated Hubbard parameters from SR- and FR-calculations do not show much differences ($<0.01$ eV) agreeing with values from our previous studies without SOCs~\cite{Yang2021PRB}. 
With FRG+$U$+$V$, we determined the equilibrium lattice constants (in~\AA) of 5.44, 5.67, 5.64 and 6.08 for Si, Ge, GaAs and GaSb, respectively, in excellent agreements with experiments~\cite{madelung2004semiconductors}. 

\begin{table}[t]
\caption{The energy gap, interband gap, and spin-orbit splitting in eV for Si, computed using FR-LDA, FRL+\textit{U}+\textit{V}, FR-GGA, and FRG+\textit{U}+\textit{V}, compared to $GW$ calculations and experiments.}
\label{tabular:Si}
\begin{ruledtabular}
\begin{tabular}{c|cccccc}
     & $E_g$ & $E^{\Gamma}_{6c-8v}$ & $\Delta^\Gamma_v$ &$\Delta^\Gamma_c$  
     &$\Delta^\text{L}_v$ & $\Delta^\text{L}_c$\\ \hline
$GW$+SOC\footnote{\label{Malone}Reference~\cite{malone2013jpcm}} 
& 1.27 & 3.28 & 0.05 & 0.04 & 0.03 & 0.02 \\ 
FR-$GW$\footnote{\label{Barker}Reference~\cite{Barker2022PRB}} 
& 1.22 & 3.22 & 0.05 & 0.04 & 0.03 & 0.01 \\
FR-LDA
& 0.45 & 2.46 & 0.05 & 0.03 & 0.03 & 0.01 \\
FRL+$U$+$V$
& 1.18 & 3.38 & 0.06 & 0.04 & 0.04 & 0.02 \\
FR-GGA
& 0.44 & 2.47 & 0.05 & 0.04 & 0.03 & 0.01 \\
FRG+$U$+$V$
& 1.20 & 3.40 & 0.06 & 0.04 & 0.04 & 0.02 \\
Exp
& 1.22\footnote{\label{cardona1}Reference~\cite{cardona2005rmp}} 
& 3.34\footnote{\label{masovic}Reference~\cite{masovic1983jpc}} & 0.044\footnote{\label{Si:madelung}Reference~\cite{madelung2004semiconductors}}  &  0.03$\sim$0.04\footref{Si:madelung} & 0.03\footref{Si:madelung} & - \\
\end{tabular}    
\end{ruledtabular}
\end{table}

\begin{table}[b]
\caption{The energy gap, interband gap, and spin-orbit splitting in eV for Ge, computed using FR-LDA, FRL+\textit{U}+\textit{V}, FR-GGA, and FRG+\textit{U}+\textit{V}, compared to $GW$ calculations and experiments.}
\label{tabular:Ge}
\begin{ruledtabular}
\begin{tabular}{c|cccccc}
        & $E_{g}$ 
        & $E^{\Gamma}_{7c-8v}$  & $\Delta^{\Gamma}_v$  & $\Delta^{\Gamma}_c$ & $\Delta^\text{L}_v$  & $\Delta^\text{L}_c$     \\ \hline
$GW$+SOC\footnote{\label{Malone}Reference~\cite{malone2013jpcm}}                       
                  & 0.54 & 0.38 & 0.32 & 0.24 & 0.20   &  0.12    \\
FR-$GW$\footnote{\label{Barker}Reference~\cite{Barker2022PRB}}
                  & 0.74 & 0.96 & 0.30 & 0.21 & 0.19   & 0.08     \\
FR-LDA               & 0.13 & 0.14 & 0.31 & 0.22 & 0.19   & 0.10    \\
FRL+$U$+$V$       & 0.53 & 0.68 & 0.40 & 0.29 & 0.25   & 0.11    \\
FR-GGA               & 0.00 & $-$0.19 & 0.30 & 0.21 & 0.19   & 0.09    \\
FRG+$U$+$V$       & 0.33 & 0.33 & 0.39 & 0.28 & 0.24   & 0.11       \\
Exp\footnote{\label{madelung}Reference~\cite{madelung2004semiconductors}}    
                  & 0.79 & 0.90 & 0.297& 0.200& 0.228& -\\
\end{tabular}
\end{ruledtabular}
\end{table}

\begin{table}[t]
\caption{The energy gap and spin-orbit splitting in eV for GaAs, computed using FR-LDA, FRL+\textit{U}+\textit{V}, FR-GGA, and FRG+\textit{U}+\textit{V}, compared to $GW$ calculations and experiments.}
\label{tabular:GaAs}
\begin{ruledtabular}
\begin{tabular}{c|ccccc}
        & $E_{g}$ 
        & $\Delta^{\Gamma}_v$  & $\Delta^{\Gamma}_c$ & $\Delta^\text{L}_v$  & $\Delta^\text{L}_c$     \\ \hline
$GW$+SOC\footnote{\label{Malone}Reference~\cite{malone2013jpcm}}                       
                  & 1.31 & 0.35 & 0.20 & 0.22 & 0.09      \\
FR-$GW$\footnote{\label{Barker}Reference~\cite{Barker2022PRB}}
                  & 1.49 & 0.34 & 0.17 & 0.21 & 0.07    \\
FR-LDA               & 0.54 & 0.32 & 0.19 & 0.20 & 0.08    \\
FRL+$U$+$V$       & 1.34 & 0.37 & 0.19 & 0.22 & 0.08    \\
FR-GGA               & 0.31 & 0.34 & 0.19 & 0.21 & 0.08    \\
FRG+$U$+$V$       & 1.03 & 0.39 & 0.19 & 0.23 & 0.08    \\
Exp       &1.57\footnote{\label{lautenschlager}Reference~\cite{lautenschlager1987prb}}&0.34\footnote{\label{GaAs:madelung}Reference~\cite{madelung2004semiconductors}}  & 0.171\footref{GaAs:madelung} & 0.22\footref{GaAs:madelung}   & 0.05\footref{GaAs:madelung}
\\
\end{tabular}
\end{ruledtabular}
\end{table}

\begin{table}[b]
\caption{The interband gap, and spin-orbit splitting in eV for GaSb, computed using FR-LDA, FRL+\textit{U}+\textit{V}, FR-GGA, and FRG+\textit{U}+\textit{V}, compared to $GW$ calculations and experiments.}
\label{tabular:GaSb}
\begin{ruledtabular}
\begin{tabular}{c|cccccc}
        & $E^{\Gamma}_{6c-8v}$ 
        & $E^{\text{L}}_{6c}-E^{\Gamma}_{8v}$  & $\Delta^{\Gamma}_v$  & $\Delta^{\Gamma}_c$ & $\Delta^\text{L}_v$  & $\Delta^\text{L}_c$     \\ \hline
$GW$+SOC\footnote{\label{Malone}Reference~\cite{malone2013jpcm}}                       
                  & 0.70 & 0.85 & 0.73 & 0.21 & 0.42   &  0.12    \\
FR-$GW$\footnote{\label{Barker}Reference~\cite{Barker2022PRB}}
                  & 0.82 & 0.78 & 0.73 & 0.20 & 0.42   & 0.09     \\
FR-LDA               & 0.12 & 0.24 & 0.74 & 0.22 & 0.42   & 0.12    \\
FRL+$U$+$V$       & 0.78 & 0.80 & 0.82 & 0.25 & 0.46   & 0.12    \\
FR-GGA               & $-$0.23 & 0.13& 0.72 & 0.22 & 0.42   & 0.12    \\
FRG+$U$+$V$       & 0.43 & 0.68 & 0.81 & 0.24 & 0.45   & 0.12    \\
Exp\footnote{\label{madelung}Reference~\cite{madelung2004semiconductors}}    
                  & 0.822 & 0.907 & 0.756& 0.213& 0.430& 0.13\\
\end{tabular}
\end{ruledtabular}
\end{table}

In Fig. \ref{fig:si}, we display the band structures of Si, Ge, GaAs and GaSb calculated with FR-LDA and FRL+\textit{U}+\textit{V}. It is immediately noticeable that all conduction bands shift up significantly, increasing their band gaps. As discussed before~\cite{Cococcioni2010JPC,Lee2020PRR}, the inter-site interactions dramatically improve the band gap values in these covalent materials, which are comparable to results from other sophisticated methods. We also note that the energetic positions of the lowest valence band obtained using FR-LDA and FRL+\textit{U}+\textit{V} are almost same to each other. Furthermore, for these semiconductors, the bandwidths of the upper valence bands increase with including $V$. This is slightly different from $GW$ calculations that increase the bandwidth of the upper valence bands only in Ge and GaSb~\cite{Barker2022PRB}.

In Tables~\ref{tabular:Si},~\ref{tabular:Ge},~\ref{tabular:GaAs} and~\ref{tabular:GaSb}, we compare values of the interband gaps, spin-orbit splittings and energy gaps of these semiconductors with $GW$ approximations and experiments in details. Following nomenclatures in Ref.~\cite{Barker2022PRB}, we use FR-$GW$ for full relativistic spinor $GW$ calculations while 
$GW$+SOC for treating SOCs as perturbations to $GW$ approximation without SOC~\cite{malone2013jpcm, Barker2022PRB}. Specifically, we compare energy gap ($E_g$), spin-orbit splittings of conduction (valence) bands at high symmetric $\Gamma(\text L)$-point ($\Delta^{\Gamma(\text{L})}_{c(v)}$), and interband gaps between conduction and valence bands labeled according to their irreducible representations~\cite{Barker2022PRB}. 

For Si in Table~\ref{tabular:Si}, the computed values using extended Hubbard interactions with SOCs show good agreements with those from FR-$GW$. Considering negligible contributions from $U$~\cite{Lee2020PRR}, we confirm again that the inter-site $V$ plays a pivotal role in computing band gaps and interband gap of $E^\Gamma_{6c-8v}$ regardless of XC functionals. Unlike the gaps, spin orbit splittings at $\Gamma$ and L-points are more or less similar with each other, implying marginal contributions of SOCs as expected. 

For Ge, GaAs and GaSb in Tables~\ref{tabular:Ge},~\ref{tabular:GaAs} and~\ref{tabular:GaSb}, the computed band gaps as well as interband gaps from FRL+$U$+$V$ show excellent agreements with values from FR-$GW$ as well as experiments. 
We also note that spin orbit splittings at high symmetric points are overestimated with extended Hubbard interactions. 
As was discussed in our previous study~\cite{Lee2020PRR}, the current formalism depends on pseudopotentials and XC functionals of DFT since the projectors are obtained from PPs. It is noticeable that, regardless of XC functionals used for DFT calculations, the contribution from the extended Hubbard interactions to energy band gaps and interband gaps are more or less similar. However, FR-GGA calculations significantly underestimate the energy gaps of these materials, if compared with those from FR-LDA, so that resulting gaps from FRG+$U$+$V$ are smaller than values from FR-$GW$, FRL+$U$+$V$ and experimental values.

\subsection{CdSe and PdO}

We show effects of extended Hubbard interactions on CdSe and PdO, all of which have significant SOCs. 
CdSe has a wurtzite structure with a large SOC of 407 meV~\cite{pedrotti1962pr}. DFT with local or semilocal XC approximations erroneously computes the ground electronic property of semiconducting PdO to be a metal~\cite{KANSARA2016ssc} so that extended Hubbard interactions to correct SIEs for states associated with the $d$ orbitals in the presence of the Pd-O bondings seem to be important. In Table~\ref{tarbular:CdSe_hub}, we summarize the self-consistent Hubbard parameters obtained using with FRG+$U$+$V$.
With these interactions for CdSe, we obtain the lattice constants of $a=4.30$ and $c=6.99$ in~\AA, agreeing with experimental values of 4.30~\AA~and 7.01~\AA~\cite{HOTJE2003jms} very well.

\begin{table}[t]
\caption{Calculated $U$ and $V$ (in eV) with FRG+$U$+$V$. $U_d$ and $U_p$ represent the on-site Hubbard parameters for $d$-orbitals in Cd and Pd, and and $p$-orbitals in Se and O, respectively.
$V_{dp}$ represents the inter-site Hubbard parameters between $d$- and $p$-orbitals of the adjacent atoms.}
\label{tarbular:CdSe_hub}
\begin{ruledtabular}
\begin{tabular}{cccc} 
                      & $U_d$    & $U_p$   & $V_{dp}$   \\ 
\hline
\multirow{1}{*}{CdSe} 

                      & 13.78 & 3.90 & 2.09  \\ 
\multirow{1}{*}{PdO} 

                      & 5.54  & 2.69 & 3.25  \\
\end{tabular}
\end{ruledtabular}
\end{table}

\begin{figure}[b]
	\begin{center}
		\includegraphics[width=1\columnwidth]{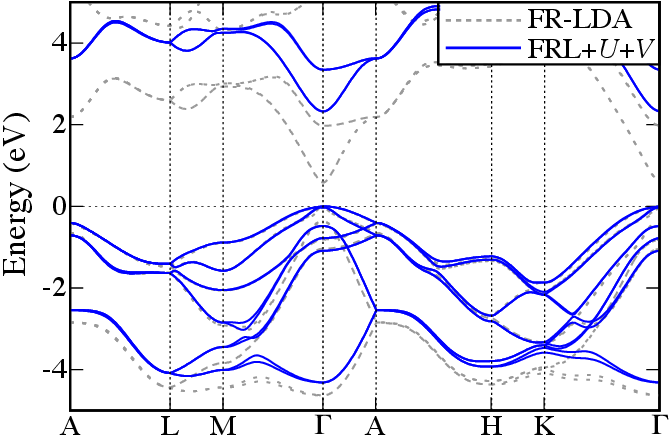}
	\end{center}
	\caption{FR-LDA and FRL+$U$+$V$~band structure of CdSe, plotted along high-symmetry lines of the Brillouin zone. The zero of the energy is at the valence band maximum.
	}
	\label{fig:CdSe}
\end{figure}

In Fig. \ref{fig:CdSe}, we plot energy bands of CdSe and confirm that the conduction bands shift up significantly with extended Hubbard interactions from those obtained by FR-LDA. The band widths of valence bands are overall slightly reduced with Hubbard interactions compared with those from FR-LDA. 
In Table \ref{tabular:CdSe}, we summarize the energy gap, spin-orbit splitting of the valence band maximum and the crystal field splitting of the valence bands at $\Gamma$ point. The energy gap from FRG+$U$+$V$ agrees with those from FR-$GW$ and experiment while the gap from FRL+$U$+$V$ is overestimated. As discussed for Ge, GaAs, and GaSb in the previous subsection, our estimations of the spin-orbit splittings are also slightly overestimated if compared with experiments. On the other hand, when the extended Hubbard correction is included, the magnitude of crystal field splitting is a little bit smaller than experiments.

\begin{table}[t]
\caption{The band gap, spin-orbit splitting, and crystal field (CF) splitting of valence band at $\Gamma$ point ($\Delta^{\Gamma}_{\text{CF},v}$) in eV for CdSe with FR-LDA, FRL+$U$+$V$, FR-GGA, FRG+$U$+$V$ compared to $GW$ calculations and experiment. }
\label{tabular:CdSe}
\begin{ruledtabular}
\begin{tabular}{c|ccc}
        & $E_{g}$& $\Delta^{\Gamma}_v$  & $\Delta^{\Gamma}_{\text{CF},v}$              \\ \hline
FR-$GW$\footnote{\label{Barker}Reference~\cite{Barker2022PRB}}                 
            & 1.85  & 0.405  & 0.026    \\
FR-LDA         & 0.58 & 0.373 & 0.037  \\
FRL+$U$+$V$ & 2.33 & 0.481 & 0.023  \\
FR-GGA         & 0.42 & 0.350 & 0.031  \\
FRG+$U$+$V$ & 1.76 & 0.481 & 0.017  \\ 
Exp.\footnote{\label{pedrotti}Reference~\cite{pedrotti1962pr}} 
            & 1.82 & 0.407 & 0.025  \\
\end{tabular}   
\end{ruledtabular}
\end{table}

\begin{table}[b]
\caption{Lattice constants and band gap of PdO computed with FR-GGA and FRG+$U$+$V$ by using FR-PPs compared with PBE and hybrid functional (HSE06) computed with SR-PPs and experiments.}
\label{tabular:PdO}
\begin{ruledtabular}
\begin{tabular}{cccc}
           & $a$ ($\text{\AA}$) & $c$ ($\text{\AA}$) & Band gap (eV)  \\ \hline
FR-GGA  & 3.04          & 5.33          & 0              \\
FRG-$U$+$V$ & 3.05          & 5.33          & 0.95           \\
PBE\footnote{\label{kansara}Scalar relativistic calculations, Reference~\cite{KANSARA2016ssc}}   & 3.14           & 5.47           & 0              \\
HSE06\footref{kansara} & 3.05           & 5.47           & 0.71           \\
Exp.     & 3.03\footnote{\label{waser}Reference~\cite{Waser1953acta}} & 5.33\footref{waser} & 0.8\footnote{\label{nilsson}Reference~\cite{Nilsson1979jpc}}, 2.13\footnote{\label{rey}Reference~\cite{rey1978jms}}, 2.67\footref{rey}    \\
\end{tabular}   
\end{ruledtabular}
\end{table}

\begin{figure}[]
	\begin{center}
		\includegraphics[width=1\columnwidth]{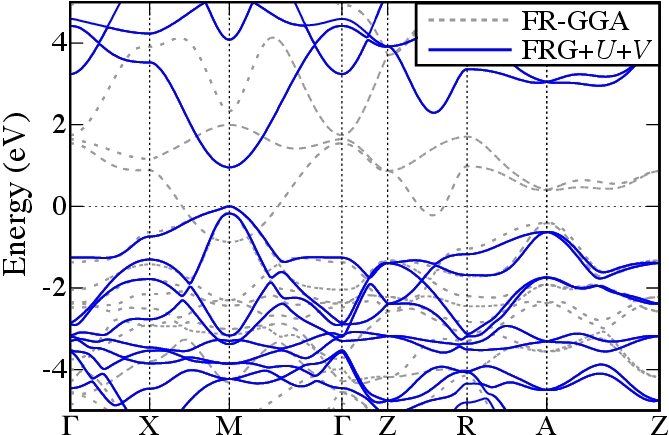}
	\end{center}
	\caption{FR-GGA and FRG+$U$+$V$~band structures of PdO, plotted along high-symmetry lines of the Brillouin zone. The Fermi energy of metallic PdO from FR-GGA and the valence band maximum from FRG+$U$+$V$ are set to zero.
	}
	\label{fig:PdO}
\end{figure}

We summarize lattice constants and energy gap of PdO in Table \ref{tabular:PdO} and plot energy bands obtained from FR-GGA and FRG+$U$+$V$ in Fig.~\ref{fig:PdO}. Like other materials discussed so far, the computed lattice constants with extended Hubbard interactions show excellent agreements with an experiment~\cite{Waser1953acta}. Furthermore, with non-negligible inter-site interactions of 3.25 eV between Pd $d$-orbitals and O $p$-orbitals in Table~\ref{tarbular:CdSe_hub}, we obtain energy gap of 0.95 eV showing qualitative agreements with an experiment~\cite{rey1978jms}, in sharp contrast to metallic ground states from FR-GGA and a scalar relativistic PBE calculation~\cite{KANSARA2016ssc}.
We also note that bandwidths of valence bands in Fig.~\ref{fig:PdO}
significantly increase from those using GGA. 

\subsection{HgTe, CuTlS$_2$, and CuTlSe$_2$}

For TIs, we first choose HgTe, CuTlS$_2$ and CuTlSe$_2$ to test our method because they are interesting materials for false-positive or false-negative assignments of topological properties depending on computational methods~\cite{Vidal2011PRB}. It has been shown that a screened hybrid functional of Heyd–Scuseria–Ernzerhof (HSE06)~\cite{Heyd2003JCP} results in inaccurate or uncertain determinations on topological characteristics of HgTe and CuTlS$_2$~\cite{Vidal2011PRB}.

Following a previous study~\cite{Vidal2011PRB}, we use the order of bands to distinguish 
three-dimensional (3D) bulk TIs from normal insulators exploiting their band inversion. The magnitude of band inversion is defined through $\Delta_i = \varepsilon_{\textit{s}} - \varepsilon_\textit{{p,d}}$ being negative at the time reversal invariant momentum. Here, $\varepsilon_{\textit{s}}$ and $\varepsilon_\textit{{p,d}}$ represent the \textit{s}-like conduction and the (\textit{p,d})-like valence band energy, respectively. 

Although DFT has been widely adopted to evaluate the topological class in various materials~\cite{chadov2010nm,lin2010nm,Xiao2010prl,Sun2010prl,Zhang2011prl,Yang2017prl}, it sometimes yields incorrect results due to the band gap underestimation~\cite{Vidal2011PRB}. Alternative computational methods such as HSE06 and the modified-Becke-Johnson (mBJ) potential are often employed to obtain improved band gaps compared with DFT-GGA or DFT-LDA~\cite{Chen2011prb,jkim2017prb,Ke2023RSC}. However, in cases of HgTe and CuTlS$_2$, HSE06 produces qualitatively different results compared to standard DFT and self-consistent $GW$ calculation (sc-$GW$)~\cite{Vidal2011PRB}. Similarly, the mBJ+$U$ approach, which combines mBJ potential with Hubbard $U$, shows positive inversion energy in CuTlS$_2$~\cite{Ke2023RSC}. Meanwhile, HSE06 and mBJ+$U$ give the same assignment of topological property for CuTlSe$_2$ if compared with other methods.

In Table \ref{tabular:TI}, we compare $\Delta_i$ from various methods. Albeit smaller magnitudes of $\Delta_i$ than those from sc-$GW$, our estimations of inversion energies for all three TIs remain to be negative, thus providing qualitatively correct assessments of their topological properties. This highlights important roles of site-dependent variations of on-site and inter-site Hubbard interactions in our method in contrast to the constant mixing of exchanges in typical hybrid approaches as was already highlighted in a previous study~\cite{Lee2020PRR} for electronic structures of the reconstructed Si(111) surface. 

\begin{table}[t]
\caption{Calculated FR-GGA and FRG+$U$+$V$ inversion energies ($\Delta_i$) compared with HSE06 and sc-$GW$ results for HgTe, CuTlS$_2$ and CuTlSe$_2$. Here we use PBE as the XC functional instead of PBEsol.}
\label{tabular:TI}
\begin{ruledtabular}
\begin{tabular}{lcccc}

                     & \multicolumn{4}{c}{$\Delta_i$ (eV)}                                                                     \\ \cline{2-5} 
  & \multicolumn{1}{c}{FR-GGA} & \multicolumn{1}{c}{sc-$GW$}\footnote{\label{Zunger}Reference~\cite{Vidal2011PRB}} & \multicolumn{1}{c}{HSE06}\footref{Zunger} & \multicolumn{1}{c}{FRG+$U$+$V$} \\ \hline
HgTe      & $-1.13$ & $-0.26$ & $\sim0$ & $-0.16$    \\
CuTlS$_2$ & $-0.72$ & $-0.41$ &  0.06   & $-0.2$    \\
CuTlSe$_2$& $-1.02$ & $-0.85$ & $-0.4$  & $-0.43$     \\ 
\end{tabular}
\end{ruledtabular}
\end{table}

\subsection{Bi$_2$Se$_3$ and Bi$_2$Te$_3$}

\begin{figure}[t]
	\begin{center}
		\includegraphics[width=1\columnwidth]{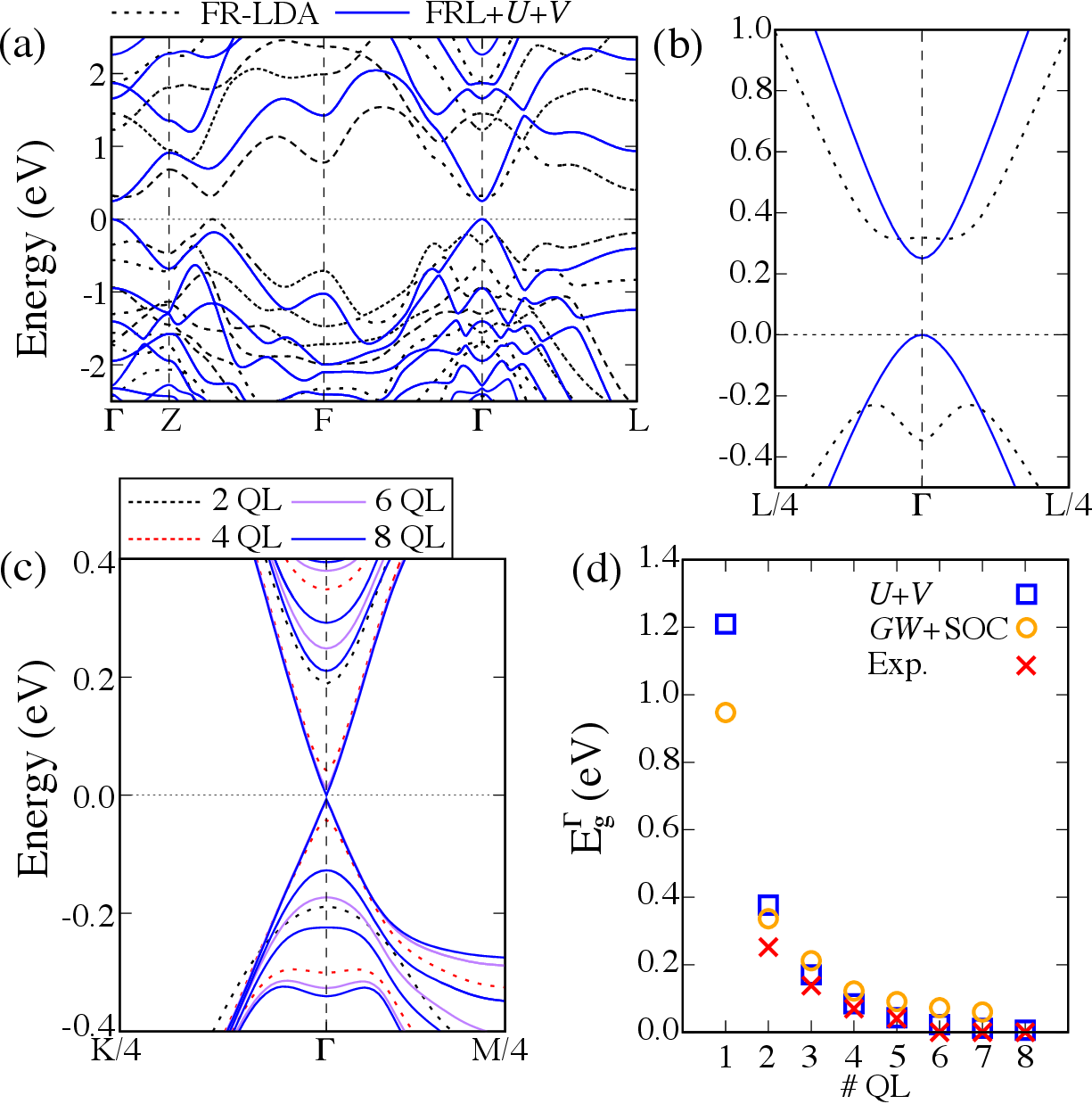}
	\end{center}
	\caption{FR-LDA and FRL+$U$+$V$~band structures of bulk Bi$_{2}$Se$_{3}$ plotted along (a) high-symmetry lines of the Brillouin zone, (b) enlarged ones near the $\Gamma$ point and (c) surface energy bands in the vicinity of the $\Gamma$ point as a function of slab thickness. (d) Band gap of topological surface bands at the $\Gamma$ point as a function of slab thickness. $GW$+SOC and experimental results are reproduced from Refs.~\cite{yazyev2012prb} and~\cite{zhang2010np}, respectively. The valence band maximum is set to zero in (a) and (b) while in (c), the Fermi energy is set to 0 eV.}
	\label{fig:Bi2Se3}
\end{figure}

\begin{figure}[t]
	\begin{center}
		\includegraphics[width=1\columnwidth]{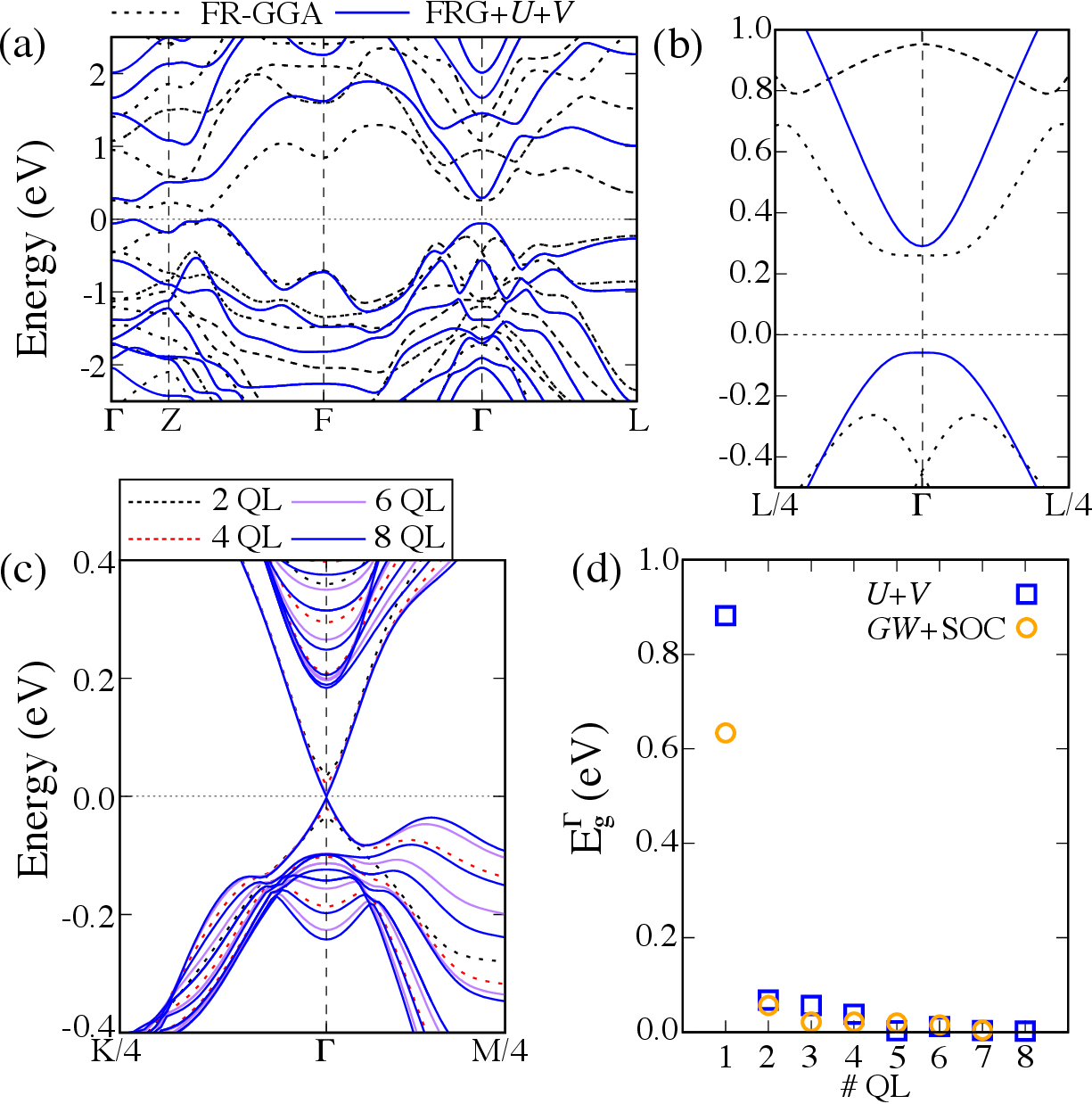}
	\end{center}
	\caption{FR-GGA and FRG+$U$+$V$~band structures of bulk Bi$_{2}$Te$_{3}$ along (a) high-symmetry lines of the Brillouin zone and (b) near the $\Gamma$ point and (c) surface bands in the vicinity of the $\Gamma$ point as a function of slab thickness. (d) Surface band gap at the $\Gamma$ point as a function of slab thickness. $GW$ calculation results are reproduced from Ref.~\cite{yazyev2012prb}. The valence band maximum is set to zero in (a) and (b) while in (c), the Fermi energy is set to 0 eV.}
	\label{fig:Bi2Te3}
\end{figure}

In Figs~\ref{fig:Bi2Se3} and~\ref{fig:Bi2Te3}, we present bulk energy bands of Bi$_2$Se$_3$ and Bi$_2$Te$_3$, both of which are well-known 3D TIs with strong SOCs~\cite{zhang2009np}. 
We used the experimental lattice parameters and atomic coordinates following previous works~\cite{nakajima1963crystal,Barker2022PRB,wyckoff1964crystal}.
Here, the strong SOC in bismuth atom causes large spin-orbit splitting in bands around the $\Gamma$ point. This brings band inversions between the valence and conduction bands across the energy gap, mixing valence and conduction states near the $\Gamma$ point, and ultimately resulting in a nontrivial Z$_2$ topological index~\cite{zhang2009np}. The size of energy gap after band inversions is proportional to the degree of mixing between the valence and conduction states. The original band gaps without SOCs are underestimated in standard DFT so that the mixing results in `camelback'-shaped DFT bands around their charge neutral energies~\cite{zhang2009np,yazyev2012prb}. This makes the DFT energy gap of 3D bulk Bi$_2$Se$_3$ and Bi$_2$Te$_3$ be indirect as shown in DFT bands in Fig.~\ref{fig:Bi2Se3}(b) and ~\ref{fig:Bi2Te3}(b).

For Bi$_2$Se$_3$, however, our FRL+$U$+$V$ calculation obtains parabolic valence and conduction bands as well as the direct band gap as shown in Figs.~\ref{fig:Bi2Se3}(b). This agrees with previous $GW$+SOC~\cite{yazyev2012prb} and FR-$GW$~\cite{Barker2022PRB} studies very well. 
The computed band gap is 0.25 eV, smaller than 0.30 eV from $GW$+SOC~\cite{yazyev2012prb} and 0.38 eV from FR-$GW$~\cite{Barker2022PRB}.  
Our computed gaps are comparable to various experimental estimations, e.g., 0.332 eV obtained from angle-resoled photoemission spectroscopy (ARPES)~\cite{Nechaev2013prb},  
0.3 eV from scanning tunneling spectroscopy (STS)~\cite{kimsh2011prl} and 0.2 eV from optical measurements~\cite{orlita2015prl}. We also estimate the effective masses from FRL+$U$+$V$ energy bands. The effective masses for the holes and electrons are 0.10 $m_e$ and 0.09 $m_e$, respectively. These values are in line with 0.19 $m_e$  (hole) and 0.14 $m_e$ (electron) from FR-$GW$ and are consistent with the 0.14 $m_e$ for both electrons and holes from magneto-optic measurements~\cite{orlita2015prl}.

Since DFT+$U$+$V$ calculation requires moderate computational cost compared to standard DFT, we test our method to calculate variation of topological surface bands (TSBs) with 
varying thickness of slab geometries. To construct supercell geometries of (111)-directional slabs with increasing thickness from one to eight QLs, atomic positions are taken from experimental data~\cite{wyckoff1964crystal} like bulk calculations. Figures \ref{fig:Bi2Se3}(c) show the evolutions of TSBs in the vicinity of the $\Gamma$ point as a function of thickness. 
It is well known that as decreasing thickness of TIs, the TSBs on opposite surfaces hybrid each other and open a gap~\cite{linder2009prb,liu2010prb,lu2010prb,zhang2010np}. Such a gap of massive TSBs in ultrathin Bi$_2$Se$_3$ decreases as its thickness increase and eventually disappear to form massless TSBs beyond a critical thickness~\cite{linder2009prb,liu2010prb,lu2010prb,zhang2010np}.

We note that within FR-LDA, TBSs of Bi$_2$Se$_3$ become to be massless beyond three or four QLs~\cite{linder2009prb,yazyev2012prb}. 
However, with extended Hubbard interactions, the gaps of massive TSBs are greatly enhanced as shown in Fig. \ref{fig:Bi2Se3}(d).
Such a trend in variation of surface band gaps agrees with $GW$+SOC calculations~\cite{yazyev2012prb} and with an experiment~\cite{zhang2010np}. Notably, for slabs thicker than three QL, our FRL+$U$+$V$ calculations showing finite gaps closely match the experimental values.
From our method, the critical thickness for gap closure in TSBs of Bi$_2$Se$_3$ is eight QLs
of which gap is below 10 meV.
For the massless dispersion,  
we estimate the group velocity of 3.6 eV$\cdot$\AA, being close to the experimental values, 3.6 eV$\cdot$\AA~\cite{kuroda2010prl} and 3.2 eV$\cdot$\AA~\cite{Zhu2011prl}. 

For Bi$_2$Te$_3$, our FRL+$U$+$V$ energy bands in Figs. \ref{fig:Bi2Te3}  (a) and (b) show a reduction of the dip in the valence bands at the $\Gamma$ point compared to the band structures from FR-GGA. Unlike Bi$_2$Se$_3$, the valence band top is more or less flat owing to a remnant corrugation of bands from mixing. The resulting indirect gaps from FR-GGA and FRG+$U$+$V$ are 0.11 and 0.25 eV, respectively, with the latter being overestimated if compared with 0.165 eV from ARPES measurement~\cite{chen2009science} and 0.17 eV from $GW$+SOC~\cite{yazyev2012prb}. For Bi$_2$Te$_3$, we use PBE as a XC functional instead of PBEsol.

For thin slabs of Bi$_2$Te$_3$, the evolution of surface band gaps are similarly dependent of computational methods as was discussed above. 
One notable difference is an oscillatory decrease of gap~\cite{linder2009prb,liu2010prb,lu2010prb} shown in Fig.~\ref{fig:Bi2Te3} (d) such that the gap of TSBs in six QLs is slightly higher than that for slabs with five QLs.
From the linear dispersion near the $\Gamma$ point in Fig. \Ref{fig:Bi2Te3} (c), the group velocities of the TSBs are estimated to be 3.0 eV$\cdot$\AA~along the $\Gamma$-M and 3.1 eV$\cdot$\AA~along the $\Gamma$-K direction. These values are close to 2.2 (2.6) eV$\cdot$\AA~along the $\Gamma$-M direction and 2.7 eV$\cdot$\AA~along the $\Gamma$-K direction from experimental measurements~\cite{chen2009science,Li2010am}. 

\section{Conclusion\label{sec:con}}

We have implemented the fully relativistic extended Hubbard functionals to calculate physical properties of materials having the spin-orbit coupling and the local and non-local Coulomb interactions simultaneously.
For on-site and inter-site interactions, we determine them self-consistently within the DFT routine by extending the ACBN0 approach~\cite{Agapito2015PRX,Lee2020PRR, Rubio2020PRB} to construct inter-site Hubbard functionals for noncollinear states. 

We applied our method to calculate atomic and electronic structures of several characteristic semiconductors and topological insulators with various strengths of SOCs. Overall, all our results for those materials show quite good agreements with fully relativistic $GW$ level calculations as well as experiments.  
Though our new method significantly improves results from  simple XC approximations in DFT, in some cases, there are noticeable dependency of computed energy gaps on XC functionals and pseudopotentials as discussed 
for the cases of III-V semiconductors. 

Considering that current formalisms without SOCs already have
provided fruitful results on phonon dispersions 
as well as electron-phonon interactions~\cite{Timrov2020PRB, Yang2021PRB,Yang2022JPC,Jang2023PRL}, our new extension 
to noncollinear DFT+$U$+$V$ is expected to be a reliable and accurate method to calculate lattice dynamical properties and electron-phonon interactions of materials with heavy atomic elements.

\begin{acknowledgements}
We thank Bradford A. Barker for providing us with fully relativistic norm conserving LDA pseudopotentials of Si, Ge, Ga, P, As, Bi, Se and Sb atoms and Myung Joon Han for discussions. Y.-W.S. was supported by KIAS individual Grant (No. CG031509). W.Y. was supported by KIAS individual Grant (No. 6P090103).
Computations were supported by the CAC of KIAS.
\end{acknowledgements}

\bibliography{soc_uv}

\end{document}